\documentclass[preprint,prc,floats,tightenlines,showpacs,superscriptaddress,nofootinbib]{revtex4-1}
\usepackage{amsfonts}
\usepackage{amssymb}
\usepackage{amsmath}
\usepackage[dvips]{graphicx,color}
\usepackage{dcolumn}
\usepackage{epsfig}
\usepackage{bm}
\usepackage{ascmac}

\usepackage{color}
\usepackage{ulem}

\newcommand{\ket}[1]{|{#1}\rangle}
\newcommand{\bra}[1]{\langle{#1}|}
\newcommand{\inp}[2]{\langle{#1}|{#2}\rangle}
\newcommand{\cg}[3]{({#1},{#2}|{#3})}

\begin{document}

\title{Toward establishing low-lying $\Lambda$ and $\Sigma$ hyperon resonances with the $\bar K + d \to \pi + Y + N$ reaction}

\author{H. Kamano}
\affiliation{KEK Theory Center, Institute of Particle and Nuclear Studies (IPNS), 
High Energy Accelerator Research Organization (KEK), Tsukuba, Ibaraki 305-0801, Japan}
\affiliation{J-PARC Branch, KEK Theory Center, IPNS, KEK, Tokai, Ibaraki 319-1106, Japan}

\author{T.-S. H. Lee}
\affiliation{Physics Division, Argonne National Laboratory, Argonne, Illinois 60439, USA}

\begin{abstract}
A  model for the $\bar K d \to \pi Y N$ reactions with $Y=\Lambda, \Sigma$ is developed,
aiming at establishing the low-lying $\Lambda$ and $\Sigma$ hyperon resonances 
through analyzing the forthcoming data from the J-PARC E31 experiment.
The off-shell amplitudes generated from the dynamical coupled-channels (DCC) model, 
which was developed in Kamano {\it et al.} [Phys.~Rev.~C {\bf 90}, 065204 (2014)], 
are used as input to the calculations of the elementary $\bar K N \to \bar K N$ and $\bar K N \to \pi Y$ 
subprocesses in the $\bar K d \to \pi Y N$ reactions.
It is shown that the cross sections for the J-PARC E31 experiment 
with a rather high incoming-$\bar{K}$ momentum, $|\vec p_{\bar K}|= 1$ GeV,
can be predicted reliably only
when the input $\bar K N \to \bar K N$ amplitudes
are generated from a $\bar KN$ model, such as the DCC model used in this investigation,
which describes the data of the $\bar K N$ reactions at energies 
far beyond the $\bar K N$ threshold.
We find that the data  of the threefold differential cross section 
$d\sigma/(dM_{\pi\Sigma}d\Omega_{p_n})$ for the $K^- d \to \pi \Sigma n$ reaction
below the $\bar K N$ threshold can be used to test the predictions
of the resonance poles associated with $\Lambda(1405)$.
We also find that the momentum dependence of the
threefold differential cross sections for the $K^- d \to \pi^- \Lambda p$ reaction
can be used to examine the existence of a low-lying $J^P=1/2^+$ $\Sigma$ resonance
with a pole mass $M_R = 1457 -i39$ MeV, 
which was found from analyzing the $K^-p$ reaction data within the employed DCC model.
\end{abstract}


\maketitle

\section{Introduction}
\label{sec:intro}

Recently, the spectroscopic study of $\Lambda$ and $\Sigma$ hyperon resonances 
with strangeness $S=-1$ (collectively referred to as $Y^*$) has made significant progress.
This advance mainly comes from using  sophisticated coupled-channels 
approaches~\cite{zhang2013,knlskp1,knlskp2,fe15} to perform comprehensive 
partial-wave analyses of the existing data of $K^- p$ reactions 
in a wide energy region from their thresholds to a rather high energy 
with the invariant mass $W=2.1$ GeV.
With this analysis, 
the systematic extraction of $Y^*$ resonances
defined by poles of the scattering amplitudes in the complex-energy plane
was accomplished.
It has been established~\cite{madrid} that the resonance poles can be identified
with the (complex-)energy eigenstates of the Hamiltonian of the underlying 
fundamental theory, which are obtained under the purely outgoing wave
boundary condition.  
Thus, the $Y^*$ resonance parameters extracted  
through the coupled-channels analyses of Refs.~\cite{zhang2013,knlskp1,knlskp2,fe15} 
have well-defined theoretical meaning, while it is often not straightforward to interpret 
the Breit-Wigner parameters listed by Particle Data Group (PDG)~\cite{pdg14}.
In addition, attempts~\cite{jlab-lqcd,ad-fvh,md} are being made to develop methods 
for relating the meson and baryon resonance poles to the lattice QCD calculations.

In this work, we consider the dynamical coupled-channels (DCC) model
developed in Ref.~\cite{knlskp1} for the meson-baryon reactions in the $S=-1$ sector. 
This model was developed by extending the theoretical framework
of Ref.~\cite{msl07}, which was originally formulated to study 
$\pi N$, $\gamma N$, $e N$, and $\nu N$ reactions
in the nucleon resonance region~\cite{jlms07,jlmss08,jklmss09,kjlms09-1,
sjklms10,kjlms09-2,ssl2,knls10,knls13,kpi2pi13,nks15,durand08,knls12,knls16}, to
include the meson-baryon channels with strangeness $S=-1$.
Within this DCC model, the $T$-matrix elements for each partial wave
can be obtained by solving a coupled integral equation~\cite{knlskp1}, 
\begin{equation}
T_{\beta,\alpha}(p_\beta,p_\alpha;W) = 
V_{\beta,\alpha}(p_\beta,p_\alpha;W)
+ \sum_{\delta} \int p^{2}d p  
V_{\beta,\delta}(p_\beta,p;W) 
G_{\delta}(p;W)
T_{\delta,\alpha}(p,p_\alpha;W) , 
\label{eq:teq}
\end{equation}
with
\begin{equation}
V_{\beta,\alpha}(p_\beta,p_\alpha;W)= 
v_{\beta,\alpha}(p_\beta,p_\alpha)
+ 
\sum_{Y^*_{0,n}}\frac{\Gamma^{\dagger}_{Y^*_{0,n},\beta}(p_\beta)
 \Gamma_{Y^*_{0,n},\alpha}(p_\alpha)} {W-M_{Y^*_{0,n}}} ,
\label{eq:veq}
\end{equation}
where $W$ is the invariant mass of the reaction;
the subscripts $\alpha$, $\beta$, and $\delta$ represent 
the five two-body channels ($\bar K N$, $\pi\Sigma$, $\pi\Lambda$, $\eta\Lambda$, and $K\Xi$) 
and the two quasi-two-body channels ($\pi\Sigma^*$ and $\bar K^* N$) that can decay into
the three-body $\pi\pi \Lambda$ and $\pi\bar K N$ channels, respectively;                                   
$p_\alpha$ is the magnitude of the momentum of channel $\alpha$ in the center-of-mass (c.m.)
frame; $G_\delta$ is the Green's function of channel $\delta$;
$M_{Y^*_{0,n}}$ is the mass of the $n$th bare excited hyperon state $Y_{0,n}^*$
included in the given partial wave;
$v_{\beta,\alpha}$ represents the hadron-exchange potentials derived from
the effective Lagrangian that respects the SU(3) flavor symmetry; 
and the bare vertex interaction 
$\Gamma_{Y^*_{0,n},\alpha}$ ($\Gamma^\dag_{Y^*_{0,n},\beta}$)
defines the $\alpha \to Y^*_{0,n}$ ($Y^*_{0,n} \to \beta$) transition.
The model parameters contained in the potential $V_{\beta,\alpha}$ were fixed by
fitting more than 17,000 data of both unpolarized and polarized observables of
the $K^- p \to \bar K N, \pi\Sigma, \pi\Lambda, \eta \Lambda, K\Xi$ reactions.
As a result, we obtained two distinct sets of the model parameters, 
referred to as Model A and Model B.  
Both models describe the existing $K^- p$ reaction data equally well 
over a wide energy range from the thresholds up to $W = 2.1$~GeV.
From Model A (Model B), 18 (20) of $Y^*$ resonances were extracted
in the energy region above the $\bar K N$ threshold and below $W = 2.1$ GeV. 
It is found that some of the extracted
low-lying $Y^*$ resonances may correspond to one- and/or two-star resonances
assigned by Particle Data Group~\cite{pdg14} or may be new resonances.
Furthermore, two $J^P=1/2^-$ $\Lambda$ resonances are found below
the $\bar K N$ threshold in both Model A and Model B,
which is similar to the results from
the chiral unitary models (see, e.g., Refs.~\cite{om,pis})
and the J\"ulich model~\cite{juelich11}.

Although a number of new and/or unestablished low-lying $Y^*$ resonances 
were found in the DCC analysis of Refs.~\cite{knlskp1,knlskp2},
their existence and pole-mass values are rather different between
Model A and Model B.
This is, of course, attributable to the fact that the existing $K^- p$ reaction data 
used in the analysis are \textit{incomplete}, as discussed in Refs.~\cite{knlskp1,knlskp2}.
In addition, there is a limitation of using the $K^- p$ reaction data 
for establishing low-lying $Y^*$ resonances because
the $K^- p$ reactions cannot directly access the energy region below the $\bar K N$
threshold, and also it is practically not easy to measure precisely the $K^- p$ reactions
in the energy region just above the $\bar K N$ threshold where
the incoming-$\bar K$ momentum becomes very low.
One of the most promising approaches to overcome this limitation
would be a combined analysis of the $K^- p$ reactions and the $K^- d \to \pi Y N$ reactions. 
This is based on the observation that 
the two-body $\pi Y$ subsystem 
in the final state of the $K^- d \to \pi Y N$ reactions
can be in the energy region below the $\bar K N$ threshold
even if the incoming-$\bar K$ momentum is rather high.

\begin{figure}[t]
\includegraphics[clip,width=0.6\textwidth]{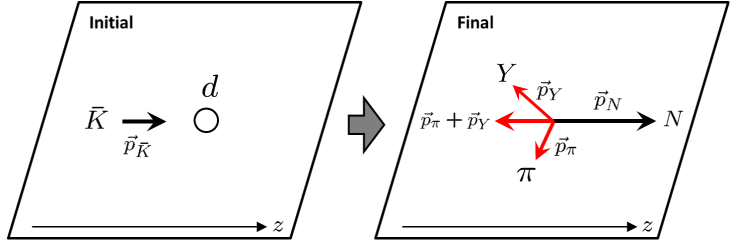}
\caption{\label{fig:kinematics}
Kinematics of the $\bar K d \to \pi Y N$ reaction considered in this work.
The outgoing $N$ (outgoing $\pi Y$ pair) momentum is in the direction (opposite direction) of the incoming-$\bar K$ momentum.
}
\end{figure}

\begin{figure}[t]
\includegraphics[clip,width=0.4\textwidth]{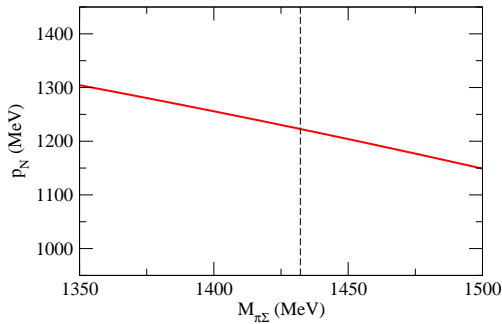}
\caption{\label{fig:kinematics-1}
The outgoing nucleon momentum $p_N\equiv |\vec p_N|$ (solid curve) 
as a function of  the kinematically allowed $\pi Y$ invariant mass $M_{\pi Y}$  
for the incoming-$\bar K$ momentum $|\vec p_{\bar K}| = 1$ GeV.
Here the case that $Y=\Sigma$ is presented.
Dashed vertical line indicates the $\pi \Sigma$ invariant mass at the $\bar K N$ threshold.
}
\end{figure}

As a first step towards accomplishing such a combined analysis of 
the $\bar K N$ and $\bar K d$ reactions, 
in this work we apply the multiple scattering theory~\cite{watson,feshbach} to
predict the differential cross sections of the $\bar K d \to \pi Y N$ reaction by using
the $\bar KN$ reaction  amplitudes generated from the DCC model of Ref.~\cite{knlskp1}.
We focus on the kinematics
that the incoming $\bar K$ has a rather high momentum of $|\vec p_{\bar K}|=1$~GeV
and the outgoing nucleon $N$ is detected at very forward angles with $\theta_{p_N}\sim 0$,
which is the same as the setup of the J-PARC E31 experiment~\cite{j-parc-e31}.
At this special parallel kinematics,
the outgoing $N$ and the outgoing $\pi Y$ pair 
are scattered back-to-back, as illustrated in Fig.~\ref{fig:kinematics},
and have almost no correlation in experimental measurements. 
In fact, as can be seen from Fig.~\ref{fig:kinematics-1},
the forward moving nucleon momenta (solid curve) 
become $|\vec p_N| > |\vec p_K|=1$~GeV
for the invariant mass of the $\pi Y$ subsystem relevant to our study (horizontal axis), 
which means that the momentum of the $\pi Y$ pair
is in an opposite direction to $\vec p_N$.
Consequently, it is the best for examining $Y^*$ resonances through 
their decays into $\pi Y$ states.
In addition, because the forward moving nucleon carries high energy-momentum,
the recoiled $\pi Y$ pair can be even below the $\bar KN$ threshold,
which is also illustrated in Fig.~\ref{fig:kinematics-1}.
We thus can make predictions for investigating 
low-lying $Y^*$ resonances, 
including the long-standing problem associated with $\Lambda(1405)$
that was also the focus of Refs.~\cite{mh,jido,ohnishi,yamagata}. 
The data from the J-PARC E31 experiment~\cite{j-parc-e31}
can then be used to  test our results. In particular, we would like to examine how the
predicted cross sections can be used to distinguish the resonance parameters extracted within 
Model A and Model B employed in our calculations.

\begin{figure}[t]
\includegraphics[clip,width=0.8\textwidth]{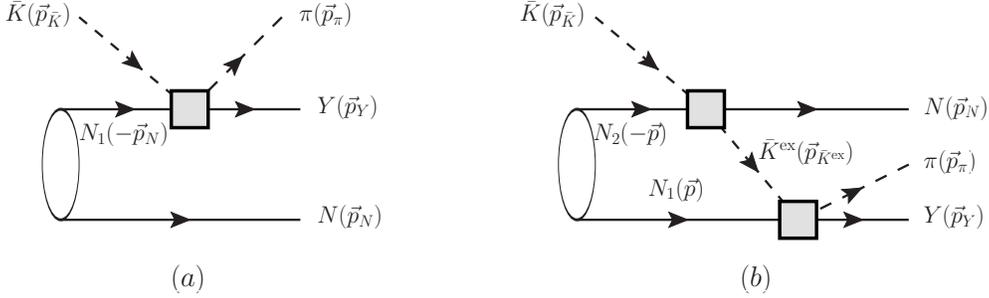}
\caption{\label{fig:reaction}
Diagrammatical representation of the $\bar K d \to \pi Y N$ reaction processes considered in this work:
(a) the impulse process; (b) the $\bar K$-exchange process.
The deuteron wave function (open circles) is taken from the one constructed with the Argonne V18
potential~\cite{dwf}, while the off-shell amplitudes describing
the meson-baryon subprocesses (solid squares)
are taken from our DCC model developed in Ref.~\cite{knlskp1}.
}
\end{figure}

Following the previous works~\cite{mh,jido,yamagata} and justified by the special
kinematics mentioned above, we assume that the scattering amplitude for $\bar K d \to \pi Y N$
includes the single-scattering (impulse) term and the $\bar K$-exchange term, as illustrated
in Fig.~\ref{fig:reaction}.
While such a perturbative approach neglects the higher-order
scattering processes in a recent calculation~\cite{ohnishi} based on 
the Alt-Grassberger-Sahdhas type of three-body scattering formulation~\cite{ags}, 
it is supported by many earlier studies of
intermediate- and high-energy reactions on deuteron; see, for example, a recent study of
$\gamma d \to \pi NN$ of Ref.~\cite{wsl15}.
Thus, it is reasonable to assume that
our results as well as the results of Refs.~\cite{mh,jido,yamagata} account for the
main features of the $\bar Kd \to \pi Y N$ reaction and can be used to
explore the feasibility of using the experiment at J-PARC to investigate
the low-lying hyperon resonances.

\begin{figure}[t]
\includegraphics[clip,width=0.75\textwidth]{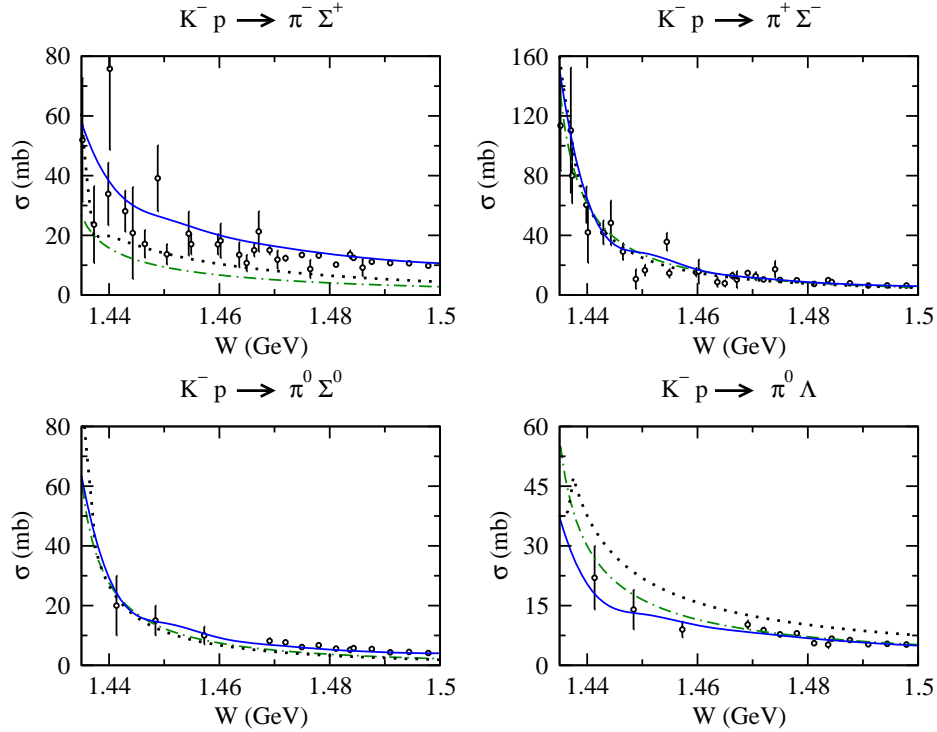}
\caption{
\label{fig:kn-tcs-low}
Total cross sections for $K^- p \to \pi Y$ reactions near the threshold.
The blue solid curves are Model B in Ref.~\cite{knlskp1},
the green dot-dashed curves are the E-dep. model in Ref.~\cite{ohnishi},
and
the black dotted curves are from the model developed in Ref.~\cite{or}
that was used for the calculation in Refs.~\cite{mh,jido,yamagata}.
}
\end{figure}

An essential difference between this work and the previous works~\cite{mh,jido,ohnishi,yamagata}
is that we employ the (off-shell) $\bar K N$ reaction amplitudes generated from the DCC model
developed in Ref.~\cite{knlskp1}.
This DCC model describes the $\bar K N$ reaction data
over a very wide energy range from the thresholds up to $W = 2.1$ GeV.
However, the models for the meson-baryon subprocesses
employed in Refs.~\cite{mh,jido,ohnishi,yamagata}
were constructed by fitting only the $K^- p$ reaction data just near the $\bar K N$ threshold.
To see how these $\bar K N$ models can be used in the calculations, 
it is instructive here to examine the kinematics of the $\bar K$-exchange mechanism illustrated
in Fig.~\ref{fig:reaction}(b).
The range of the invariant mass of the outgoing $\pi Y$ system  ($M_{\pi Y}$)
we are interested in is $m_\pi + m_Y \leq M_{\pi Y}\lesssim 1.5$ GeV, where $m_\pi$ ($m_Y$) is the mass of $\pi$ ($Y$).
Thus the $\bar K^{\rm ex} N_1 \to \pi Y$ amplitudes used for calculating the $\bar K$-exchange 
mechanism must be generated from models which can reproduce well the data near the $\bar KN$ threshold.
As seen in Fig.~\ref{fig:kn-tcs-low},  
the models used in Refs.~\cite{mh,jido,ohnishi} and the DCC models 
employed in our calculations are all valid for this calculation in
the invariant mass $M_{\pi Y} $ covered by the J-PARC E31 experiment  
shown in Fig.~\ref{fig:kinematics-1}.

\begin{figure}[t]
\includegraphics[clip,width=0.45\textwidth]{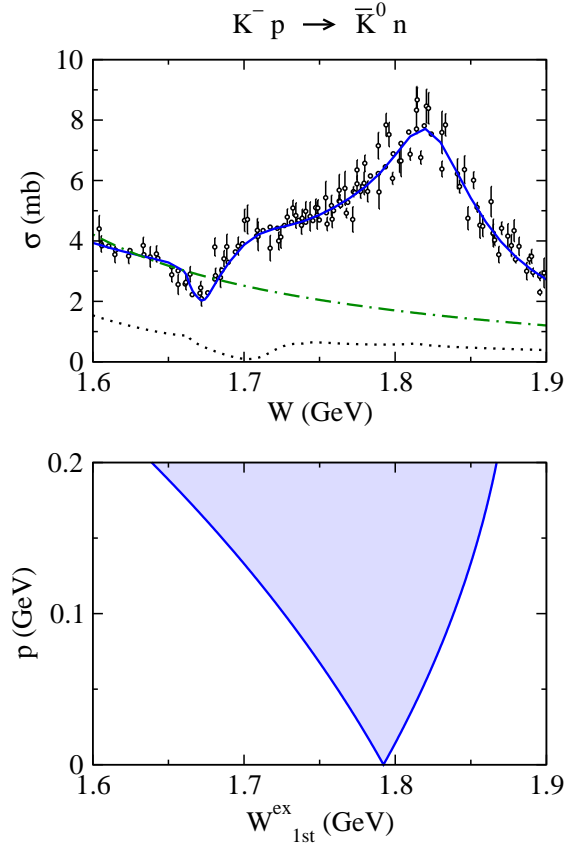}
\caption{
\label{fig:kn-tcs-high}
(Top) Total cross section for $K^- p \to \bar K^0 n$ 
in the energy region relevant to the $\bar K N_2 \to \bar K^{\rm ex} N$ subprocess
in the $\bar K$-exchange process [Fig.~\ref{fig:reaction}(b)].
The blue solid curve is Model B in Ref.~\cite{knlskp1},
the green dot-dashed curve is the E-dep. model in Ref.~\cite{ohnishi}, and 
the black dotted curve is from the model developed in Ref.~\cite{or}
that was used for the calculation in Refs.~\cite{mh,jido,yamagata}.
(Bottom) Allowed ranges of the invariant mass $W^{\rm ex}_{\rm 1st}$
for the $\bar K N_2 \to \bar K^{\rm ex} N$ subprocess as $p\equiv|-\vec p|$ is varied. 
Here the incoming-$\bar K$ momentum and the scattering angle of outgoing $N$ are fixed as
$|\vec p_{\bar K}|=1$~GeV and $\theta_{p_N} = 0$, respectively.
}
\end{figure}

The situation is very different for the calculations of $\bar K N_2 \to \bar K^{\rm ex} N$ amplitudes
in Fig.~\ref{fig:reaction}(b).
In the bottom panel of Fig.~\ref{fig:kn-tcs-high}, 
we show the ranges of the invariant mass ($W^{\rm ex}_{\rm 1st}$) 
of the $\bar K N_2 \to \bar K^{\rm ex} N$ subprocess,
which can be formed from the incoming-$\bar K$ momentum $|\vec p_K| = 1$~GeV,
the scattering angle of outgoing $N$ $\theta_{p_N}=0$, 
and the momentum of initial nucleon $N_2$ with $|-\vec p| < 0.2$~GeV
within which the deuteron wave function is large.
We see that 
for a rather high incoming-$\bar K$ momentum with $|\vec p_K| = 1$~GeV,
the allowed ranges for $W^{\rm ex}_{\rm 1st}$
are in the well above the $\bar K N$ threshold region.
In the top panel of Fig.~\ref{fig:kn-tcs-high}, we see that only
the DCC model can describe the data in the whole range.
Thus, the models used in Refs.~\cite{mh,jido,ohnishi} have large uncertainties
in calculating the $\bar K N_2 \to \bar K^{\rm ex} N$ amplitudes for predicting 
$\bar K d \rightarrow \pi Y N$ at $|\vec p _{\bar K}|=1 $ GeV
to compare with the data from  the J-PARC E31 experiment~\cite{j-parc-e31}.
In this work, we will also discuss how these uncertainties 
associated with the $\bar K N_2 \to \bar K^{\rm ex} N$ amplitudes
affect the resulting $\bar K d \rightarrow  \pi Y N$ reactions cross sections.

In Sec.~\ref{sec:formula}, we first give the notations 
for kinematical variables and the cross section formulas
necessary for the presentation of this work.
We then  give the formula for calculating the impulse and
$\bar K$-exchange  amplitudes of the $\bar K d \rightarrow \pi YN$ reactions.
The predicted results for the $\bar K d \to \pi Y N$ reaction from 
our model are presented in Sec.~\ref{sec:results}. 
The comparisons with the results from using the $S$-wave $\bar K N$ models are also given there.
A summary and the prospect for future works are given in Sec.~\ref{sec:summary}.

\section{Formulation}
\label{sec:formula}

In this section, we present the formulas for the calculations of the differential
cross sections for $\bar K + d \to \pi + Y + N$ that can be used to compare
with the data from the J-PARC E31 experiment.

\subsection{Kinematics and cross sections}

We perform calculations  
in the laboratory (LAB) frame in which the incoming $\bar K$ is in the
quantization $z$ direction and the outgoing $N$ is on the $x$-$z$ plane. 
The momenta for the $\bar K + d \to \pi + Y + N$ reaction, 
denoted as $p_a$ ($a=\bar K, d, \pi, Y, N$), can then be written as
\begin{eqnarray}
p_{\bar K} &=& (E_{\bar K}(\vec p_{\bar K}), 0, 0, |\vec p_{\bar K}|),
\label{eq:k1}\\
p_d &=& (m_d, \vec 0),
\\
p_\pi &=& (E_\pi(\vec p_\pi), \vec p_\pi),
\\
p_Y &=& (E_Y(\vec p_Y), \vec p_Y),
\\
p_N &=& (E_N(\vec p_N), |\vec p_N|\sin\theta_{p_N}, 0, |\vec p_N|\cos\theta_{p_N} ),
\label{eq:k5}
\end{eqnarray}
where $E_a(\vec p_a) = (m_a^2 +\vec p_a^2)^{1/2}$ is the relativistic energy for a particle $a$ with mass $m_a$ and momentum $\vec p_a$.
It is convenient to introduce
the momentum $\vec q_{\pi}$ of the outgoing $\pi$ in the center-of-mass (c.m.)
frame of the final $\pi Y$ subsystem. 
For a given invariant mass $M_{\pi Y}$ of
the $\pi Y$ subsystem, the magnitude of $\vec q_{\pi}$ is given by
\begin{equation}
|\vec q_\pi| = \frac{1}{2M_{\pi Y}}\sqrt{\lambda(M_{\pi Y}^2,m_\pi^2,m_Y^2)},
\end{equation}
where $\lambda(a,b,c)$ is the K\"allen function defined by $\lambda(a,b,c)=a^2+b^2+c^2-2ab-2bc-2ac$.
For given $M_{\pi Y}$ and $\cos\theta_{p_N}$, $|\vec p_{N}|$ is obtained by solving
$E_{\bar K}(\vec p_{\bar K})+m_d= E_{N}(\vec p_N) + E_{\pi Y}$ where
$E_{\pi Y} = \sqrt{M_{\pi Y}^2+\vec P_{\pi Y}^2}$ and 
$\vec P_{\pi Y} \equiv \vec p_\pi + p_Y = \vec p_{\bar K} - \vec p_N$.
The momenta $\vec{p}_\pi$ for the outgoing $\pi$ and $\vec{p}_Y$ 
for the outgoing $Y$ can then be given by
\begin{equation}
\vec p_{\pi} = \vec q_{\pi}  + \frac{\vec P_{\pi Y}}{M_{\pi Y}}
\left[
\frac{\vec P_{\pi Y} \cdot \vec q_\pi}{E_{\pi Y}+ M_{\pi Y}} + E_\pi(\vec q_\pi)
\right],
\end{equation}
\begin{equation}
\vec p_{Y} =-\vec q_{\pi}  + \frac{\vec P_{\pi Y}}{M_{\pi Y}}
\left[
-\frac{\vec P_{\pi Y} \cdot \vec q_\pi}{E_{\pi Y}+ M_{\pi Y}} + E_Y (\vec q_\pi)
\right],
\end{equation}
With the above formulas, the kinematical variables [Eqs.~(\ref{eq:k1})-(\ref{eq:k5})]
are completely fixed by the incoming-$\bar K$ momentum $\vec p_{\bar K}$, 
the solid angle $\Omega_{p_N}=( \theta_{p_N}, \phi_{p_N}\equiv 0)$ of the outgoing $N$ 
on the $x$-$z$ plane, 
the solid angle $\Omega_{q_\pi}=(\theta_{q_\pi}, \phi_{q_\pi})$ of the outgoing $\pi$ 
in the $\pi Y$ c.m. frame, and the $\pi Y$ invariant mass $M_{\pi Y}$.

With the normalization $\inp{\vec p'}{\vec p} = \delta(\vec p' - \vec p)$ 
for the plane-wave one-particle state,
the unpolarized differential cross sections investigated in this work are given by
\begin{equation}
\frac{d\sigma}{dM_{\pi Y} d\Omega_{p_N}} = 
\int d\Omega_{q_\pi}
\frac{d\sigma}{dM_{\pi Y} d\Omega_{p_N}d\Omega_{q_\pi}} ,
\label{eq:dcs3fold}
\end{equation}
\begin{eqnarray}
\frac{d\sigma}{dM_{\pi Y} d\Omega_{p_N}d\Omega_{q_\pi}} 
&=& 
(2\pi)^4 
\frac{E_{\bar K}(\vec p_{\bar K})}{|\vec p_{\bar K}|}
\frac{E_{\pi}(\vec p_\pi)E_{Y}(\vec p_Y)E_{N}(\vec p_N)|\vec q_\pi||\vec p_{N}|^2}
{\left| \left[E_{\bar K}(\vec p_{\bar K})+m_d \right]|\vec p_N|- E_N(\vec p_N) |\vec p_{\bar K}|\cos\theta_{p_N}\right|}
\nonumber\\
&&
\times
\frac{1}{(2J_d+1)}\sum_{\textrm{spins}}|T_{\pi Y N,\bar K d}|^2 ,
\label{eq:dcs5fold}
\end{eqnarray}
where $d\Omega_p = d\phi_{p} d\cos\theta_p$; $J_d = 1$ is the spin of the deuteron;
and $T_{\pi Y N, \bar K d}$ is the $T$-matrix element for the $\bar K d \to \pi Y N$ reaction.

\subsection{Model for $\bar K d \to \pi Y N$  reaction}

As discussed in Sec.~\ref{sec:intro}, the cross section for the $\bar K d \to \pi YN$ reaction 
will be calculated from the mechanisms illustrated in Fig.~\ref{fig:reaction}.
The $T$-matrix element $T_{\pi Y N,\bar K d}$ appearing in Eq.~(\ref{eq:dcs5fold})
is given as a sum of contributions from the impulse ($T_{\pi Y N, \bar K d}^{\rm imp}$) and 
$\bar K$-exchange  ($T_{\pi Y N, \bar K d}^{\bar K\textrm{-ex}}$) processes:
\begin{equation}
T_{\pi Y N,\bar K d} = T_{\pi Y N, \bar K d}^{\rm imp} + T_{\pi Y N, \bar K d}^{\bar K\textrm{-ex}}.
\end{equation}

The $T$-matrix element for the impulse process [Fig.~\ref{fig:reaction}(a)] is given by
\begin{eqnarray}
T_{\pi Y N, \bar K d}^{\rm imp} &=&
\sqrt{2}
\bra{\pi(\vec p_\pi, I^z_{\pi});Y(\vec p_Y, S^z_Y,I^z_Y); N(\vec p_N, S^z_N,I^z_N)}
t_{\pi Y, \bar K N_1}
\ket{\Psi_d^{(M_d)}; \bar K (\vec p_{\bar K}, I^z_{\bar K})}
\nonumber\\
&=&
\sqrt{2}
\sum_{S^z_{N_1}}
T_{\pi(I^z_\pi) Y(S^z_Y,I^z_Y),\bar K (I^z_{\bar K}) N_1 (S^z_{N_1},-I^z_N)}
(\vec p_\pi,\vec p_Y; \vec p_{\bar K},-\vec p_N;W^{\rm imp})
\nonumber\\
&&
\qquad
\times
\Psi_d^{(M_d)}(-\vec p_N,S^z_{N_1}, -I^z_{N}; \vec p_N, S^z_{N}, I^z_{N}),
\label{eq:t-imp}
\end{eqnarray}
where $I^z_a$ ($S^z_a$) is the quantum number for the $z$ component of 
the isospin $I_a$ (the spin $S_a$) of the particle $a$; and $M_d$ is that of the deuteron spin.
The factor $\sqrt{2}$ comes from the antisymmetry property of the deuteron wave function
given by the following standard form:
\begin{eqnarray}
\Psi_d^{(M_d)}(\vec p, m_{s1}, m_{t1}; -\vec p, m_{s2}, m_{t2})
&=&
\cg{\frac{1}{2}m_{t1}}{\frac{1}{2}m_{t2}}{00}
\nonumber\\
&&\times
\sum_{LM_LM_s}
\cg{LM_{L}}{1M_{s}}{1M_d}
\cg{\frac{1}{2}m_{s1}}{\frac{1}{2}m_{s2}}{1M_s}
\nonumber\\
&&\times
Y_{LM_L}(\hat p) R_L(|\vec p|),
\end{eqnarray}
Here $\cg{l_1m_1}{l_2m_2}{lm}$ is the Clebsch-Gordan coefficient for $l_1\otimes l_2\to l$;
$Y_{LM}(\hat p)$ is the spherical harmonics; and $R_L(|\vec p|)$ is the radial wave function.
The radial wave function is normalized as
\begin{eqnarray}
\sum_{L=0,2}\int^\infty_0 p^2\,dp\,|R_L(p)|^2 =1 .
\end{eqnarray}
In this work, the radial wave function, $R_L(|\vec p|)$ with $L=0,2$,
is taken from Ref.~\cite{dwf}.

The half-off-shell $\bar K N_1 \to \pi Y$ scattering in Eq.~(\ref{eq:t-imp}) 
can be related to the one in its c.m. frame by
\begin{multline}
T_{\pi(I^z_\pi) Y(S^z_Y,I^z_Y),\bar K (I^z_{\bar K}) N_1(S^z_{N_1},-I^z_N)}
(\vec p_\pi,\vec p_Y; \vec p_{\bar K},-\vec p_N;W^{\rm imp})
=
\\
\qquad
\sqrt{
\frac{E_\pi(\vec q_\pi)E_Y(-\vec q_\pi)E_{\bar K}(\vec q_{\bar K})E_N(-\vec q_{\bar K})}
{E_\pi(\vec p_\pi)E_Y(\vec p_Y)E_{\bar K}(\vec p_{\bar K})E_N(-\vec p_N)}
}
T^{\text{c.m.}}_{\pi(I^z_\pi) Y(S^z_Y,I^z_Y),\bar K(I^z_{\bar K}) N_1 (S^z_{N_1},-I^z_N)}
(\vec q_\pi,-\vec q_\pi; \vec q_{\bar K},-\vec q_{\bar K}; W^{\rm imp}),
\label{eq:ele-amp1}
\end{multline}
where $\vec q_{\bar K}$ is the momentum of the incoming $\bar K$ in the c.m. frame of 
the final $\pi Y$ system; the Lorentz-boost factor appears in the right-hand side\footnote{
Strictly speaking, the Wigner rotations also take place for the particle spins
through the Lorentz boost.
However, those are omitted here because those do not affect the unpolarized 
differential cross sections considered in this work.
}; and
the invariant mass $W^{\rm imp}$ for the $\bar K N_1 \to \pi Y$ subprocess 
is defined by 
\begin{equation}
W^{\rm imp} = M_{\pi Y}.
\end{equation}
Furthermore, the partial-wave expansion of the amplitude in the c.m. frame is expressed as
\begin{multline}
T^{\text{c.m.}}_{\pi(I^z_\pi) Y(S^z_Y,I^z_Y),\bar K(I^z_{\bar K}) N_1(S^z_{N_1},-I^z_N)}
(\vec q_\pi,-\vec q_\pi; \vec q_{\bar K},-\vec q_{\bar K}; W^{\rm imp})
=\\
\sum_{JLJ^zL_f^zL_i^z}\sum_{II^z} Y_{LL^z_f}(\hat q_f)Y^*_{LL^z_i}(\hat q_i)
\cg{LL_f^z}{S_Y S_Y^z}{JJ^z}
\cg{LL_i^z}{S_{N_1} S_{N_1}^z}{JJ^z}
\\
\times
\cg{I_\pi I_\pi^z}{I_Y I_Y^z}{II^z}
\cg{I_{\bar K} I_{\bar K}^z}{I_{N_1} -I_N^z}{II^z}
\
T_{\pi Y, \bar K N_1}^{(IJL)} (q_\pi,q_{\bar K}; W^{\rm imp}).
\label{eq:ele-amp2}
\end{multline}
As already mentioned, in this work we take the partial-wave amplitudes
$T_{\pi Y, \bar K N_1}^{(IJL)} (q_\pi,q_{\bar K}; W^{\rm imp})$ from the DCC model developed in Ref.~\cite{knlskp1}.

For the $\bar K$-exchange process [Fig.~\ref{fig:reaction}(b)], 
the corresponding $T$-matrix element
is expressed as
\begin{eqnarray}
T_{\pi Y N,\bar K d}^{\bar K\textrm{-ex}} &=&
\sqrt{2}
\bra{\pi(\vec p_\pi, I^z_{\pi});Y (\vec p_Y, S^z_Y,I^z_Y);N (\vec p_N, S^z_N,I^z_N)}
\nonumber\\
&&
\qquad
\times
\hat t_{\pi Y,\bar K^{\rm ex} N_1}
\hat G_{\bar K^{\rm ex} N N_1} 
\hat t_{\bar K^{\rm ex} N,\bar K N_2}
\ket{\Psi_d^{(M_d)}; \bar K(\vec p_{\bar K}, I^z_{\bar K})}
\nonumber\\
&=&
\sum_{S_{N_1}^z S_{N_2}^z}
\sum_{I^z_{\bar K^{\rm ex}}I_{N_1}^zI_{N_2}^z}
\int d\vec p_{\bar K^{\rm ex}}
\nonumber\\
&&
\qquad
\times
T_{\pi(I^z_\pi) Y(S^z_Y,I^z_Y),\bar K^{\rm ex}(I^z_{\bar K^{\rm ex}}) N_1(S_{N_1}^z,I_{N_1}^z)}
(\vec p_\pi,\vec p_Y; \vec p_{\bar K^{\rm ex}},\vec p;W^{\rm ex}_{\rm 2nd})
\nonumber\\
&& 
\qquad
\times
\frac{1}{E-E_{\bar K^{\rm ex}}(\vec p_{\bar K^{\rm ex}})-E_N(\vec p_N)-E_{N_1}(\vec p) +i\varepsilon}
\nonumber\\
&&
\qquad
\times
T_{\bar K^{\rm ex}(I^z_{\bar K^{\rm ex}}) N(S^z_N,I^z_N), \bar K (I^z_{\bar K}) N_2 (S_{N_2}^z,I_{N_2}^z)}
(\vec p_{\bar K^{\rm ex}},\vec p_N; \vec p_{\bar K} , -\vec p;W^{\rm ex}_{\rm 1st})
\nonumber\\
&&
\qquad
\times
\Psi_d^{(M_d)}(\vec p, S_{N_1}^z, I_{N_1}^z; -\vec p, S_{N_2}^z, I_{N_2}^z),
\label{eq:mex-amp}
\end{eqnarray}
where $\vec p = \vec p_\pi + \vec p_Y - \vec p_{\bar K^{\rm ex}} = \vec p_{\bar K} - \vec p_N - \vec p_{\bar K^{\rm ex}}$;
and $E$ is the total scattering energy in the LAB frame.
$W^{\rm ex}_{\rm 1st}$ and $W^{\rm ex}_{\rm 2nd}$ are respectively the invariant mass for 
the $\bar K N_2 \to \bar K^{\rm ex} N$ and $\bar K^{\rm ex} N_1 \to \pi Y$ subprocesses
that describe the first and second meson-baryon interaction vertices 
[solid squares in Fig.~\ref{fig:reaction}(b)]
in the $\bar K$-exchange process.
The explicit form of $W^{\rm ex}_{\rm 1st}$ and $W^{\rm ex}_{\rm 2nd}$ are given by
\begin{equation}
W^{\rm ex}_{\rm 1st} = \sqrt{[E_{\bar K}(\vec p_{\bar K}) + m_d - E_{N_1}(\vec p)]^2 - (\vec p_N + \vec p_{\bar K^{\rm ex}})^2} ,
\end{equation}
\begin{equation} 
W^{\rm ex}_{\rm 2nd} = M_{\pi Y} .
\end{equation}
Again, the off-shell plane-wave amplitude for 
the $\bar K N_2 \to \bar K^{\rm ex} N$ and $\bar K^{\rm ex} N_1 \to \pi Y$ subprocesses
are constructed with the partial-wave amplitudes generated from the DCC model~\cite{knlskp1}
in a way similar to Eqs.~(\ref{eq:ele-amp1}) and~(\ref{eq:ele-amp2}).

\section{Results and Discussion}
\label{sec:results}

With the model described  in the previous section, we can use
Eqs.~(\ref{eq:dcs3fold}) and~(\ref{eq:dcs5fold}) to calculate
the differential cross sections for the $K^- d \to \pi Y N$ reactions. 
We will first present our predictions
for using  the forthcoming data from the J-PARC E31 experiment to examine
the low-lying $Y^*$ resonances 
that were extracted~\cite{knlskp2} from the two DCC models, Model A and Model B, 
of Ref.~\cite{knlskp1}. 
We then discuss the differences between our results
with those given in Refs.~\cite{mh,jido,ohnishi}.

\subsection{Predictions for J-PARC E31 experiment}

To make predictions for  the J-PARC E31 experiment, we consider the kinematics that
the momentum of the incoming $K^-$ is set as $|\vec p_{K^-}| = 1$~GeV 
and the momentum of the outgoing $N$ is chosen to be 
in the $K^-$ direction with $\theta_{p_N}=0$.
We perform calculations using 
the $\bar K N \to \bar K N$ and $\bar K N \to \pi Y$ amplitudes generated from
both of the DCC models (Model A and Model B) constructed in Ref.~\cite{knlskp1}.
The predicted $K^-d$ results are denoted as Model A and Model B accordingly.

First of all, we observe that 
the impulse process [Fig.~\ref{fig:reaction}(a)] gives negligible contribution 
at the considered kinematics with $|\vec p_{K^-}| = 1$ GeV and $\theta_{p_N} = 0$,
and the cross sections are completely dominated by 
the $\bar K$-exchange process [Fig.~\ref{fig:reaction}(b)].
This is expected because the impulse amplitude~(\ref{eq:t-imp}) contains 
the deuteron wave function $\Psi_d(-\vec{p}_N, \vec{p}_N)$, 
which becomes very small in the considered kinematics where
the momentum $\vec p_N$ is very high, $|\vec p_N| \sim 1.2$~GeV, as indicated in Fig.~\ref{fig:kinematics-1}.
Therefore, in the following, our discussions are focused on the $\bar K$-exchange process.

\begin{figure}[t]
\includegraphics[clip,width=\textwidth]{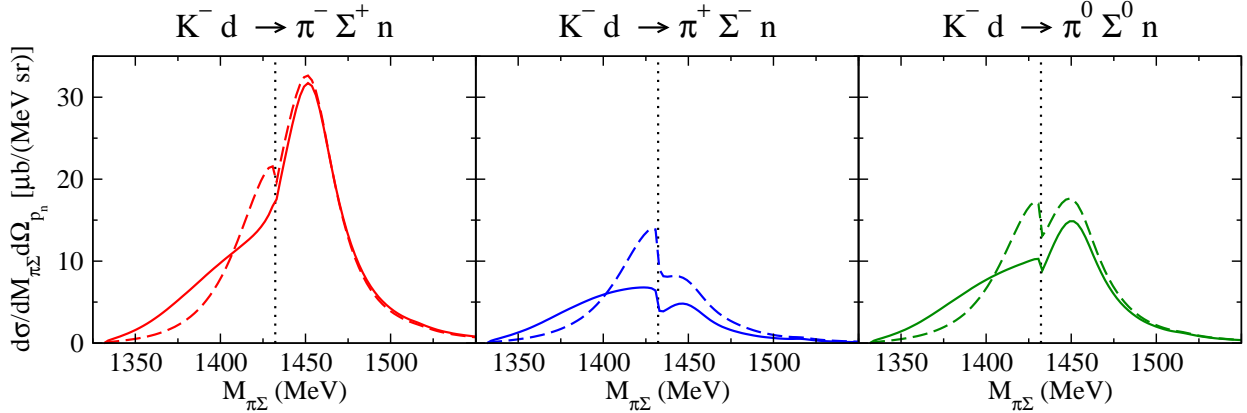}
\caption{\label{fig:piSfull}
Threefold differential cross section
$d\sigma/(dM_{\pi \Sigma} d\Omega_{p_n})$ for the
$K^-d \to \pi \Sigma n$ reactions at $|\vec p_{K^-}| = 1$ GeV and $\theta_{p_n} = 0$.
Solid curves (dashed curves) are the full results for which 
the off-shell partial-wave amplitudes of Model A (Model B) of our DCC model~\cite{knlskp1}
are used for the two-body meson-baryon subprocesses.
Dotted vertical lines indicate the $\pi \Sigma$ invariant mass at the $\bar K N$ threshold.
}
\end{figure}

Figure~\ref{fig:piSfull} shows
the predicted threefold differential cross section $d\sigma/(dM_{\pi \Sigma}d\Omega_n)$
for the $K^- d \to \pi \Sigma n$ reactions.
There are two noticeable features.
First, there is a significant enhancement of the cross section 
at $M_{\pi \Sigma} \sim 1.45$ GeV. 
Second, a varying structure, partly attributable to the cusp from the opening of
the $\bar K N$ channel, appears in the considered $M_{\pi \Sigma}$ region,
and its shape depends on
the model and the charge state of the final $\pi \Sigma$ system.
We analyze their origins in the following.

The enhancement of the cross section in Fig.~\ref{fig:piSfull}
at $M_{\pi \Sigma} \sim 1.45$ GeV is mainly attributable to the fact that 
the meson-baryon amplitudes are, in general, the largest at the on-shell kinematics and
the deuteron wave function $\Psi_d (\vec{p},-\vec{p})$ is the largest at $|\vec{p}|=0$.
At $M_{\pi \Sigma} \sim $ 1.45 GeV, all of the meson-baryon subprocesses and three-body
propagator in the $\bar K$-exchange process become almost on-shell
when the momenta of the nucleons inside the deuteron are near $|\vec{p}|=0$ in
the integrand of Eq.~(\ref{eq:mex-amp}).
Thus, the magnitude of $\bar K$-exchange amplitude $|T^{\bar K\textrm{-ex}}_{\pi Y N, \bar K d}|$
gets a large enhancement at $M_{\pi \Sigma} \sim 1.45$ GeV.
This is similar to what was discussed in Ref.~\cite{ohnishi}.
In fact, we confirm that the enhancement disappears if we omit the contribution
from the $|\vec{p}| < 0.2$ GeV region in the loop integration in Eq.~(\ref{eq:mex-amp}).

\begin{figure}[t]
\includegraphics[clip,width=0.6\textwidth]{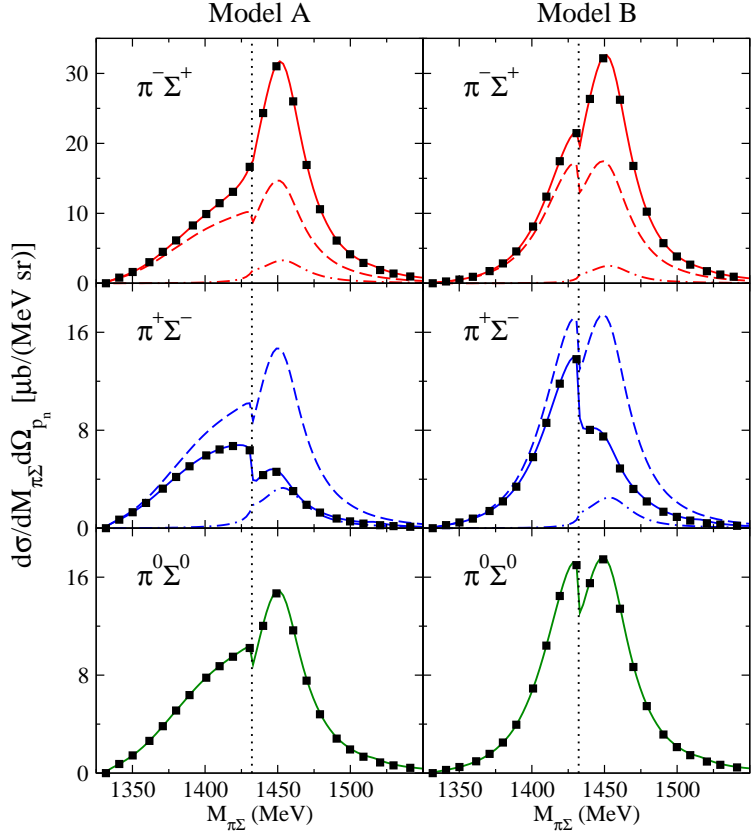}
\caption{\label{fig:piS-bd}
Threefold differential cross section $d\sigma/(dM_{\pi \Sigma} d\Omega_{p_n})$ 
for the $K^-d \to \pi \Sigma n$ reactions with $|\vec p_{K^-}| = 1$ GeV and $\theta_{p_n} = 0$.
Top, middle, and bottom panels are the results for
$K^-d \to \pi^- \Sigma^+ n$,
$K^-d \to \pi^+ \Sigma^- n$, and
$K^-d \to \pi^0 \Sigma^0 n$, respectively.
The results from Model A (Model B) are presented in left panels (right panels).
Each of the curves and points is
the full results (solid curves), and
the results in which only 
the $S$-wave amplitude (solid squares), 
the $S_{01}$ amplitude (dashed curves), or 
the $S_{11}$ amplitude (dashed-dotted curves) 
is included in $\bar K^{\rm ex} N_1 \to \pi \Sigma$ of the $\bar K$-exchange process.
Dotted vertical lines indicate the $\pi \Sigma$ invariant mass at the $\bar K N$ threshold.
}
\end{figure}

We now examine the varying structure of $d\sigma/(dM_{\pi\Sigma}d\Omega_{p_n})$ 
in Fig.~\ref{fig:piSfull}.
For this purpose, we first observe in Fig.~\ref{fig:piS-bd} that 
the results (solid squares) 
from keeping only the $S$ wave of the $\bar K^{\rm ex} N_1 \to \pi \Sigma$ amplitude 
agree almost perfectly with the full results (solid curves). 
This indicates that the $\bar K^{\rm ex} N_1 \to \pi \Sigma$ subprocess
is completely dominated by the $S$-wave amplitudes in the considered kinematics.
We note that this explains why a peak owing to 
the $\Lambda(1520)3/2^-$ resonance does not appear at $M_{\pi \Sigma}\sim 1.52$~GeV
in contrast to the case of the $K^- p$ reactions.
In the same figure, we also show the contributions
from $S_{01}$ (dashed curves) and $S_{11}$ (dashed-dotted curves) partial 
waves\footnote{
The partial wave of the two-body $\bar K +N \to M(0^-)+B(\frac{1}{2}^+)$ reactions is denoted as $L_{I2J}$, 
which means that the partial wave has
a total angular momentum $J$, a total isospin $I$, and a parity $P = (-)^L$.
}
of the $\bar K^{\rm ex} N_1 \to \pi \Sigma$ subprocess.
Clearly, the main contributions to the full results (solid curves)
are from the $S_{01}$ wave that show the clear cusp structure near the $\bar K N$ threshold. 
However, their interference with the $S_{11}$
wave is significant and is constructive (destructive)
for the $\pi^- \Sigma^+$ ($\pi^+ \Sigma^-$) production reactions.
Such interference is absent
for the $\pi^0 \Sigma^0$ production reaction, because only
the $S_{01}$ wave of the $\bar K^{\rm ex} N_1 \to \pi \Sigma$ 
subprocess can contribute to the cross section.

\begin{figure}[t]
\includegraphics[clip,width=0.6\textwidth]{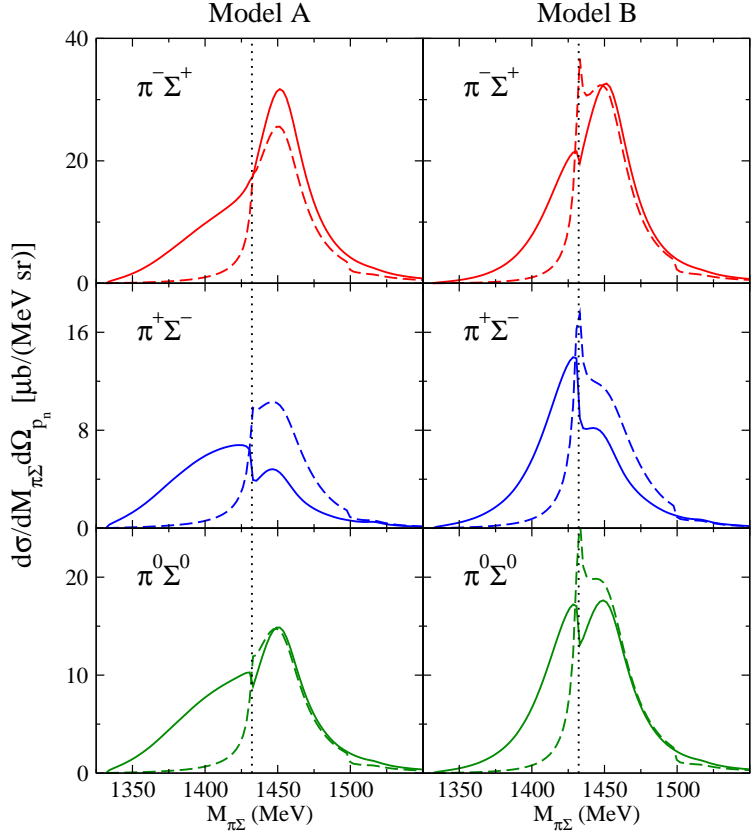}
\caption{\label{fig:piS-nr}
Threefold differential cross section $d\sigma/(dM_{\pi \Sigma} d\Omega_{p_n})$
for the $K^-d \to \pi \Sigma n$ reactions with $|\vec p_{K^-}| = 1$ GeV and $\theta_{p_n} = 0$.
Top, middle, and bottom panels are the results for
$K^-d \to \pi^- \Sigma^+ n$,
$K^-d \to \pi^+ \Sigma^- n$, and
$K^-d \to \pi^0 \Sigma^0 n$, respectively.
The results from Model A (Model B) are presented in left panels (right panels).
Solid curves are the full results, while dashed curves are the same as solid curves,
except that only the nonresonant contribution is included for the $S_{01}$ amplitude of 
the $\bar K^{\rm ex} N_1 \to \pi \Sigma$ subprocess.
Dotted vertical lines indicate the $\pi \Sigma$ invariant mass at the $\bar K N$ threshold.
}
\end{figure}

\begin{figure}[t]
\includegraphics[clip,width=0.4\textwidth]{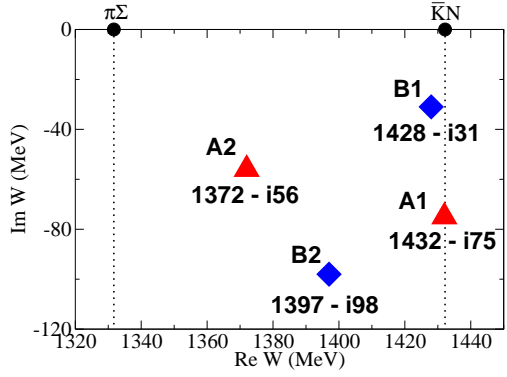}
\caption{\label{fig:s01-pole}
Pole positions of $S_{01}$ ($J^P=1/2^-$) $\Lambda$ resonances located 
below the $\bar K N$ threshold~\cite{knlskp2},
which were extracted within the DCC models developed in Ref.~\cite{knlskp1}.
Red triangles (blue diamonds) are the resonance pole positions obtained from Model A (Model B).
}
\end{figure}

\begin{table}[t]
\caption{\label{tab:residue}
The product of coupling strengths $g_{\pi\Sigma Y^*}g_{\bar K NY^*}$
at pole positions for $J^P=1/2^-$ $\Lambda$ resonances located below the $\bar K N$ threshold.
The pole mass $M_R$ is presented as $(\textrm{Re}(M_R),-\textrm{Im}(M_R))$,
and $g_{\pi\Sigma Y^*}g_{\bar K NY^*}=|g_{\pi\Sigma Y^*}g_{\bar K NY^*}| e^{i\phi}$ is presented as
($|g_{\pi\Sigma Y^*}g_{\bar K NY^*}|$, $\phi$).
The product $g_{\pi\Sigma Y^*}g_{\bar K NY^*}$ is defined as the residue of 
the $T$-matrix element $T_{\pi \Sigma,\bar K N}$
at the resonance pole position.
}
\begin{ruledtabular}
\begin{tabular}{cccc}
  & Pole mass $M_R$ (MeV) & $g_{\pi\Sigma Y^*}g_{\bar K NY^*}$ (MeV$^{-1}$, deg.)&
$\left|g_{\pi\Sigma Y^*}g_{\bar K NY^*}/\textrm{Im}(M_R)\right|^2$ (MeV$^{-4}$)\\
\hline
A1&$(1432,75)$& ($15.42 \times 10^{-4}$, $ 170$)& $4.23\times 10^{-10}$\\
B1&$(1428,31)$& ($ 7.94 \times 10^{-4}$, $ 102$)& $6.56\times 10^{-10}$\\
\\
A2&$(1372,56)$& ($21.54 \times 10^{-4}$, $- 24$)& $14.79\times 10^{-10}$\\
B2&$(1397,98)$& ($13.87 \times 10^{-4}$, $- 56$)& $2.00\times 10^{-10}$
\end{tabular}
\end{ruledtabular}
\end{table}

We next examine how the characteristic differences between Model A and Model B
in the shape of the cross sections below the $\bar K N$ threshold 
(compare solid and dashed curves in Fig.~\ref{fig:piSfull})
can be related to resonances in the $S_{01}$ partial wave of 
the $\bar K^{\rm ex} N_1 \to \pi \Sigma$ subprocess. 
For this purpose, we first observe in Fig.~\ref{fig:piS-nr} that 
the cross sections become very small below the $\bar K N$ threshold
if we take into account only the nonresonant contribution for the $S_{01}$ wave 
of $\bar K^{\rm ex} N_1 \to \pi \Sigma$. 
With this observation, we expect that $S_{01}$ ($J^P=1/2^-$) $\Lambda$ resonances 
are actually the main contribution of the cross sections below the $\bar K N$ threshold
and are the origin of the difference in its shape between Model A and Model B.
As mentioned in Sec.~\ref{sec:intro}, our DCC analysis of the $K^- p$ reactions~\cite{knlskp1}
predicts two $S_{01}$ ($J^P=1/2^-$) $\Lambda$ resonances below 
the $\bar K N$ threshold in both Model A and Model B~\cite{knlskp2}, 
as shown in Fig.~\ref{fig:s01-pole}.
Here, the higher mass pole (A1 and B1) would correspond to the $\Lambda(1405)$ resonance,
while another $\Lambda$ resonance with lower mass (A2 and B2) is similar to 
what was obtained in the chiral unitary models (see, e.g., Refs.~\cite{om,pis}) 
and the J\"ulich model~\cite{juelich11}.
Although both Model A and Model B find two $\Lambda$ resonances, 
their pole positions are rather different.
One can see from Fig.~\ref{fig:s01-pole} that the pole A1 (B2) has a larger 
imaginary part than the pole B1 (A2) and is far away from the real energy axis.
In addition, the products of their coupling strengths
to the $\pi \Sigma$ and $\bar K N$ channels, $g_{\pi \Sigma Y^*} \times g_{\bar K N Y^*}$,
are rather different, as seen in Table~\ref{tab:residue}.
The contribution of a resonance with complex mass $M_R$ 
in the $\bar K^{\rm ex} N_1 \to \pi \Sigma$ subprocess
to the $\bar K$-exchange amplitude $T^{\bar K\textrm{-ex}}_{\pi \Sigma n,K^- d}$ 
can be schematically expressed at $M_{\pi \Sigma} = {\rm Re}(M_R)$ as
\begin{eqnarray}
T^{\bar K\textrm{-ex}}_{\pi \Sigma n,K^- d} 
&\sim& 
\left[ 
F(M_{\pi \Sigma}) \times \frac{g_{\pi \Sigma Y^*}g_{\bar K N Y^*}}{M_{\pi \Sigma} -M_R} + \cdots 
\right]_{M_{\pi \Sigma} = {\rm Re}(M_R)} 
\nonumber\\
&=& 
F\bm{(}{\rm Re}(M_R)\bm{)}
\times
\frac{g_{\pi \Sigma Y^*}g_{\bar K N Y^*}}{i{\rm Im}(M_R)} + \cdots,
\end{eqnarray}
where $F(M_{\pi \Sigma})$ is a regular function of $M_{\pi \Sigma}$ and is expected
not to be much different between Model A and Model B.
The value of $|g_{\pi \Sigma Y^*} g_{\bar K N Y^*} / {\rm Im}(M_R)|^2$
can therefore be used to measure the effect of a resonance on the cross section.
In the third column of 
Table~\ref{tab:residue}, we see that 
$|g_{\pi \Sigma Y^*} g_{\bar K N Y^*} / {\rm Im}(M_R)|^2$ of 
the resonance B1 is larger than that of A1. 
Thus B1 has larger effects than A1 on the cross sections near the $\bar KN$ threshold, 
as can be seen 
from clear peaks in the cross sections at $M_{\pi \Sigma}\sim 1.42$~GeV
that appear only in Model B.
At lower energy, the cross sections are influenced by the
second resonances A2 and B2. 
From Table~\ref{tab:residue}, we see that 
$|g_{\pi \Sigma Y^*} g_{\bar K N Y^*} / {\rm Im}(M_R)|^2$ of 
the resonance A2 is much larger than that of B2. 
This explains why the cross sections at 
$M_{\pi\Sigma} \lesssim 1.4$ GeV in Model A are larger than those in Model B.

\begin{figure}[t]
\includegraphics[clip,width=0.6\textwidth]{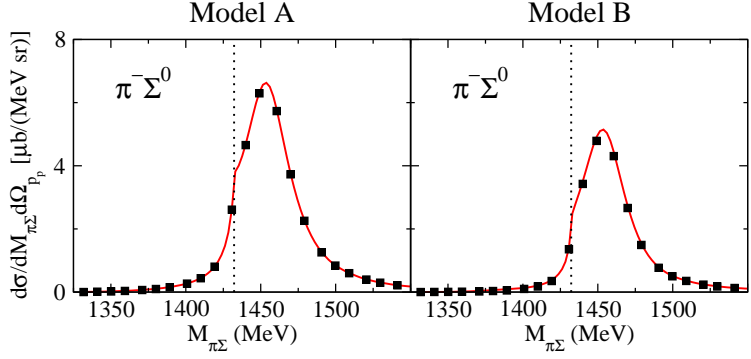}
\caption{\label{fig:pi-S0}
Threefold differential cross section
$d\sigma/(dM_{\pi \Sigma} d\Omega_{p_p})$ for the
$K^-d \to \pi^- \Sigma^0 p$ reaction at $|\vec p_{K^-}| = 1$ GeV and $\theta_{p_p} = 0$.
The results from Model A (Model B) are presented in the left panel (right panel).
Solid curves are the full results, while solid squares are
the results in which only the $S_{11}$ amplitude is included for
$\bar K^{\rm ex} N_1 \to \pi \Sigma$ of the $\bar K$-exchange process.
Dotted vertical lines indicate the $\pi \Sigma$ invariant mass at the $\bar K N$ threshold.
}
\end{figure}

\begin{figure}[t]
\includegraphics[clip,width=0.6\textwidth]{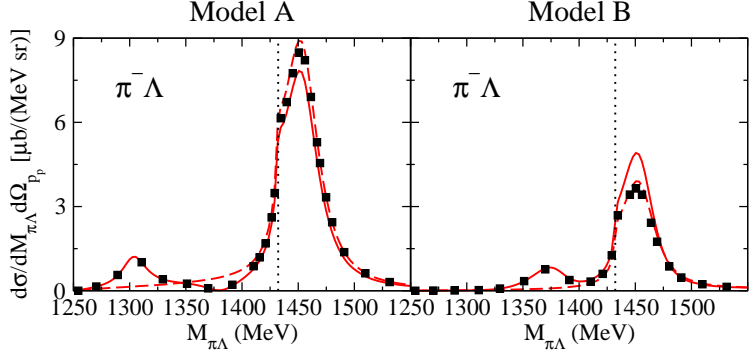}
\caption{\label{fig:pi-L}
Threefold differential cross section
$d\sigma/(dM_{\pi \Lambda} d\Omega_{p_p})$ for the
$K^-d \to \pi^- \Lambda p$ reaction at $|\vec p_{K^-}| = 1$ GeV and $\theta_{p_p} = 0$.
The results from Model A (Model B) are presented in the left panel (right panel).
Each of curves and points is
the full results (solid curves), and
the results in which only 
the $S_{11}$ amplitude (dashed curves) or
the $S_{11}$ and $P_{13}$ amplitudes (solid squares)
are included for $\bar K^{\rm ex} N_1 \to \pi \Lambda$ of the $\bar K$-exchange process.
Dotted vertical lines indicate the $\pi \Lambda$ invariant mass at the $\bar K N$ threshold.
}
\end{figure}

We now turn to presenting the predicted cross sections
for $K^-d \to \pi^- \Sigma^0 p$ and $K^-d \to \pi^- \Lambda p$ at
the same kinematics $|\vec p_{\bar K}|=1$ GeV and $\theta_{p_p} = 0$.
Because the $\pi^- \Sigma^0$ and $\pi^- \Lambda$ states contain only the isospin $I=1$ component,
these reactions will be useful for investigating the low-lying $\Sigma$ resonances.
It is noted that the data for such reactions can also be obtained 
by extending the measurements of the the J-PARC E31 experiment~\cite{noumi}.
Similar to the results for the $K^- d \to \pi \Sigma n$ reactions presented above, 
we find that 
(a) the impulse process gives negligible contribution to the cross sections for both 
$K^-d \to \pi^- \Sigma^0 p$ and $K^-d \to \pi^- \Lambda p$ and
(b) the characteristic enhancement appears at $M_{\pi Y}\sim 1.45$ GeV,
as seen in Figs.~\ref{fig:pi-S0} and~\ref{fig:pi-L}.

For $K^-d \to \pi^- \Sigma^0 p$, we find that 
the $\bar K^{\rm ex} N_1 \to \pi \Sigma$ subprocess is completely dominated
by the $S_{11}$ amplitude. 
This is shown in Fig.~\ref{fig:pi-S0}.
We see that the results (solid squares)  from the calculations keeping only the
$S_{11}$ amplitude of the $\bar K^{\rm ex} N_1 \to \pi \Sigma$ subprocess
agree almost perfectly with the results (solid curves) from the calculations including all partial waves.
The cross section becomes very small below the $\bar K N$ threshold, and
this would be because no resonance exists in the $S_{11}$ wave in the corresponding energy region.
It is found that Model B shows 
the cross section $\sim 20$ \% smaller than Model A at its maximum ($M_{\pi \Sigma}\sim 1.45$ GeV).
Because the on-shell $S_{11}$ amplitudes for the $\bar K^{\rm ex} N_1 \to \pi \Sigma$ subprocess 
are not much different between the two models at $M_{\pi \Sigma}\sim 1.45$ GeV~\cite{knlskp1}, 
the difference in the magnitude of the $K^-d \to \pi^- \Sigma^0 p$ cross section 
might partly come from that in the off-shell behavior 
of the $\bar K^{\rm ex} N_1 \to \pi \Sigma$ subprocess.

The predicted differential cross sections for the $K^- d \to \pi^- \Lambda p$ reaction are given in Fig.~\ref{fig:pi-L}.
By comparing the solid curves and the solid squares, it is clear that 
the $S_{11}$ and $P_{13}$ waves of the $\bar K^{\rm ex} N_1 \to \pi \Lambda $ subprocess completely dominate 
the cross section in the region below the $\bar K N$ threshold.
A resonance corresponding to $\Sigma(1385)3/2^+$ in the $P_{13}$ wave was identified in both Model A and Model B.
For Model B (the right panel of Fig.~\ref{fig:pi-L}), 
there is a peak at $M_{\pi \Lambda}\sim 1.38$ GeV, 
where the contribution from the $S_{11}$ amplitude is very weak.
However, we find that in Model A 
the $S_{11}$-wave contribution and the $P_{13}$-wave contribution
from $\Sigma(1385)3/2^+$ are comparable and interfere destructively, and,
as a result, a dip is produced at $M_{\pi \Lambda}\sim 1.38$ GeV.
We find that Model A has another $P_{13}$ resonance with 
lower mass than $\Sigma(1385)3/2^+$. 
This is the origin of the peak 
at $M_{\pi \Lambda}\sim 1.3$ GeV in the  left panel of Fig.~\ref{fig:pi-L}.
These kinds of visible differences between Model A and Model B can occur
below the $\bar K N$ threshold, because at present our DCC models for the $\bar K N$ reactions
have been constructed by fitting only to the $K^- p$ reaction data.
We expect that such a different behavior of the two-body subprocesses
below the $\bar K N$ threshold, which cannot be directly 
constrained by the $\bar K N$ reaction data, needs to be judged by the data of $\bar K d$ reactions.
The upcoming data from the J-PARC E31 experiment are thus highly desirable 
to improve our DCC models in the $S=-1$ sector.

We also see in Fig.~\ref{fig:pi-L} that above the $\bar K N$ threshold,
the $P_{13}$ wave of the $\bar K^{\rm ex} N_1 \to \pi \Lambda $ subprocess is negligible 
and the main contribution to the cross section comes from the $S_{11}$ wave.
However, the behavior of the $S_{11}$ partial-wave amplitudes for $\bar K N \to \pi \Lambda $ is rather different
between Model A and Model B at $W \lesssim 1.7$ GeV (see Fig.~27 in Ref.~\cite{knlskp1}),
and this is the origin of the 
the sizable difference in the magnitude of the cross section above the $\bar K N$ threshold.
For Model A (left panel), the difference between the solid and dashed curves
is quite small, and hence the cross section above the $\bar K N$ threshold
is almost completely dominated by the $S_{11}$ wave.
However, this difference is about 30 $\%$ for Model B (right panel) and 
is found to come from a $P_{11}$ ($J^P=1/2^+$) $\Sigma$ resonance with pole mass 
$M_R = 1457 -i39$~MeV~\cite{knlskp2}.
This resonance might correspond to the one-star $\Sigma(1480)$ 
resonance assigned by PDG~\cite{pdg14}.
At present this resonance was found only in Model B, and this is why 
the contribution of the $P_{11}$ wave is negligible in
the $K^-d \to \pi^- \Lambda p$ cross section for  Model A.

\begin{figure}[t]
\includegraphics[clip,width=0.7\textwidth]{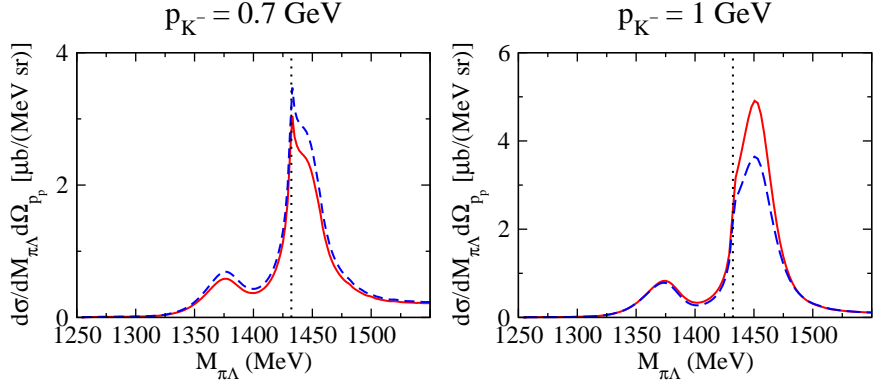}
\caption{\label{fig:3dcs-pdep}
Threefold differential cross section
$d\sigma/(dM_{\pi \Lambda} d\Omega_{p_p})$ for the
$K^-d \to \pi^- \Lambda p$ reaction at $\theta_{p_p} = 0$, computed with Model B.
The left (right) panel is the result at $|\vec p_{\bar K}| = 0.7$ GeV ($|\vec p_{\bar K}| = 1$ GeV).
Solid curves are the full results, while dashed curves are the results
in which the $P_{11}$ amplitude
for $\bar K^{\rm ex} N_1 \to \pi \Lambda$ in the $\bar K$-exchange process
is turned off.
Dotted vertical lines indicate the $\pi \Lambda$ invariant mass at the $\bar K N$ threshold.
}
\end{figure}

The above result suggests that the $K^-d \to \pi^- \Lambda p$ cross section 
may provide a useful constraint for judging this unestablished low-lying $\Sigma$ resonance
with spin-parity $J^P=1/2^+$.
To investigate this, we examine the threefold differential cross sections at different
values of the incoming-$\bar K$ momentum.
In Fig.~\ref{fig:3dcs-pdep}, we present $d\sigma/(dM_{\pi\Lambda}d\Omega_{p_p})$ at 
$|\vec p_{\bar K}|=1$ and 0.7~GeV.
We find that the interference pattern in the cross section changes as $|\vec p_{\bar K}|$ changes.
For the cross section at $|\vec p_{\bar K}|= 1$ GeV, 
the contribution from the $P_{11}$ wave of the $\bar K^{\rm ex} N_1 \to \pi \Lambda$ subprocess 
shows a constructive interference with the other contributions,
while at $|\vec p_{\bar K}|= 0.7$ GeV, it shows a destructive interference.
This visible difference of the interference pattern originating from the $P_{11}$ wave 
of the $\bar K^{\rm ex} N_1 \to \pi \Lambda$ subprocess 
will provide critical information for judging the unestablished $J^P=1/2^+$ $\Sigma$ resonance.
Therefore, it is highly desirable to measure 
the $K^-d \to \pi^- \Lambda p$ cross section 
for several $|\vec p_{\bar K}|$ values.

\subsection{Comparison with the results from the $S$-wave $\bar KN$ models}

\begin{figure}[t]
\includegraphics[clip,width=\textwidth]{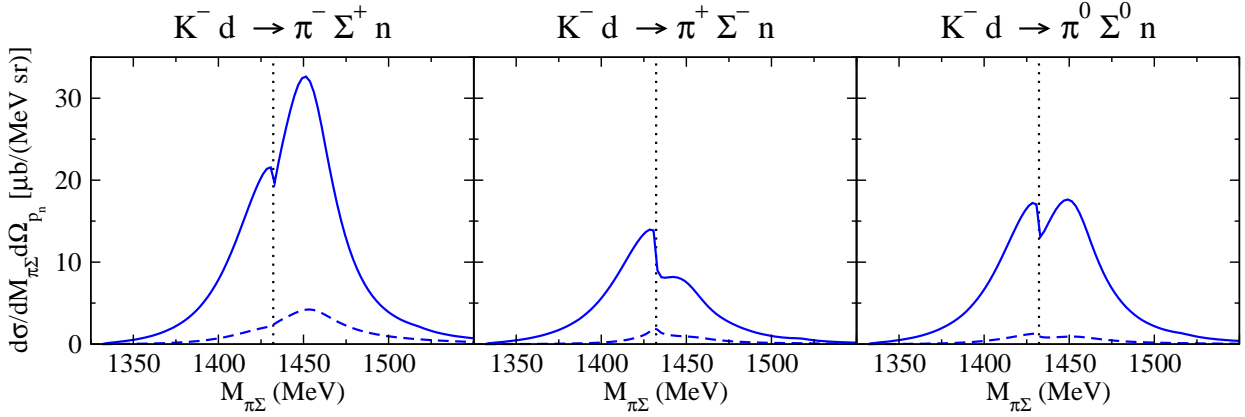}
\caption{\label{fig:piSfull-s}
Threefold differential cross section
$d\sigma/(dM_{\pi \Sigma} d\Omega_{p_n})$ for the
$K^-d \to \pi \Sigma n$ reactions at $|\vec p_{K^-}| = 1$ GeV and $\theta_{p_n} = 0$.
Solid curves represent the full result, while dashed curves 
represent the results in which only the $S$-wave amplitudes are
included for $\bar K N_2 \to \bar K^{\rm ex} N$ of 
the $\bar K$-exchange process.
Dotted vertical lines indicate the $\pi \Sigma$ invariant mass at the $\bar K N$ threshold.
}
\end{figure}

The differential cross sections at $|\vec p_K|=1$ GeV are also predicted in  Ref.~\cite{ohnishi}.
We first note that our predicted cross sections 
shown in Fig.~\ref{fig:piSfull} are much  
larger than those given in Fig.~12 of Ref.~\cite{ohnishi}. 
We find that it is mainly attributable to the large difference between the amplitudes
used in the calculations of $\bar K N_2 \to \bar K^{\rm ex} N$ 
in the $\bar K$-exchange process [Fig.~\ref{fig:reaction}(b)], 
where the incoming $\bar K$ has a large momentum. 
As seen in Fig.~\ref{fig:kn-tcs-high},
the $S$-wave $\bar K N$ model used in Ref.~\cite{ohnishi} underestimates 
the $\bar K N \to \bar K N$ cross section greatly in the invariant-mass region 
around $W = 1.8$ GeV, which is
covered in the loop integration of Eq.~(\ref{eq:mex-amp}) over the momentum of the nucleon in the
deuteron. 
In such a high-$W$ region far beyond the $\bar K N$ threshold,
it is necessary to include the higher partial-wave contributions.
This can be understood from Fig.~\ref{fig:kn-tcs-high}, where we
compare  the $K^- p \rightarrow \bar{K}^0 n$
cross sections from our DCC model and the two $S$-wave models of Refs.~\cite{ohnishi,or}.
If we keep only the $S$-wave part of the amplitude in our calculation, our
results (solid curve) in Fig.~\ref{fig:kn-tcs-high} are actually reduced to the values
close to the results (dot-dashed and dotted curves)  of the two $S$-wave models. 
Accordingly, we see in Fig.~\ref{fig:piSfull-s} that the magnitude of $d\sigma/(dM_{\pi\Sigma}d\Omega_{p_n})$ 
for the $K^- d \to \pi \Sigma n$ reactions 
are drastically reduced if we include only the $S$-wave amplitudes 
for $\bar K N_2 \to \bar K^{\rm ex} N$ in the $\bar K$-exchange process.
This result indicates that the use of appropriate amplitudes 
that reproduce the $\bar K N$ reactions up to a very high energy 
is inevitable for obtaining the $K^- d$ reaction cross sections 
that are comparable with the experimental data.
The same argument would also apply to the other studies of 
the $K^- d$ reaction~\cite{mh,jido,yamagata},
where the amplitudes for the meson-baryon subprocesses are 
obtained by fitting only to the near-threshold data of $\bar K N$ reactions.
It is noted that the higher-order scattering processes were also taken into account in Ref.~\cite{ohnishi}.
By performing calculations using their $S$-wave $\bar K N$ model, however, 
we confirm that in the considered kinematics 
their results are nearly saturated by the impulse and $\bar K$-exchange processes
and the higher-order effects seem subdominant.
Therefore, the use of appropriate $\bar K N$ scattering amplitudes, which 
can make the $K^- d$ reaction cross sections order(s) of magnitude larger, seems more important than the higher-order effects.

\begin{figure}[t]
\includegraphics[clip,width=\textwidth]{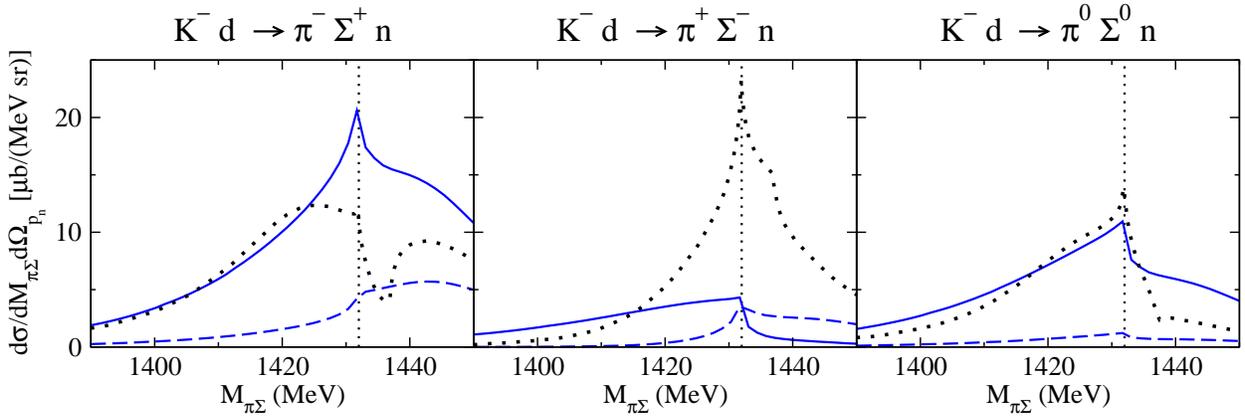}
\caption{
\label{fig:600mev}
Threefold differential cross section
$d\sigma/(dM_{\pi \Sigma} d\Omega_{p_n})$ for the
$K^-d \to \pi \Sigma n$ reactions at $|\vec p_{K^-}| = 0.6$ GeV and $\theta_{p_n} = 0$.
Solid curves are the full results from our Model B,
while dashed curves are the results from Model B in which only the 
the $S$-wave amplitudes are included for all meson-baryon subprocesses.
Dotted curves are the results in Ref.~\cite{mh}, where
the $S$-wave $\bar K N$ model developed in Ref.~\cite{or} are used for
calculating the meson-baryon subprocesses.
Dotted vertical lines indicate the $\pi \Sigma$ invariant mass at the $\bar K N$ threshold.
}
\end{figure}

\begin{figure}[t]
\includegraphics[clip,width=0.45\textwidth]{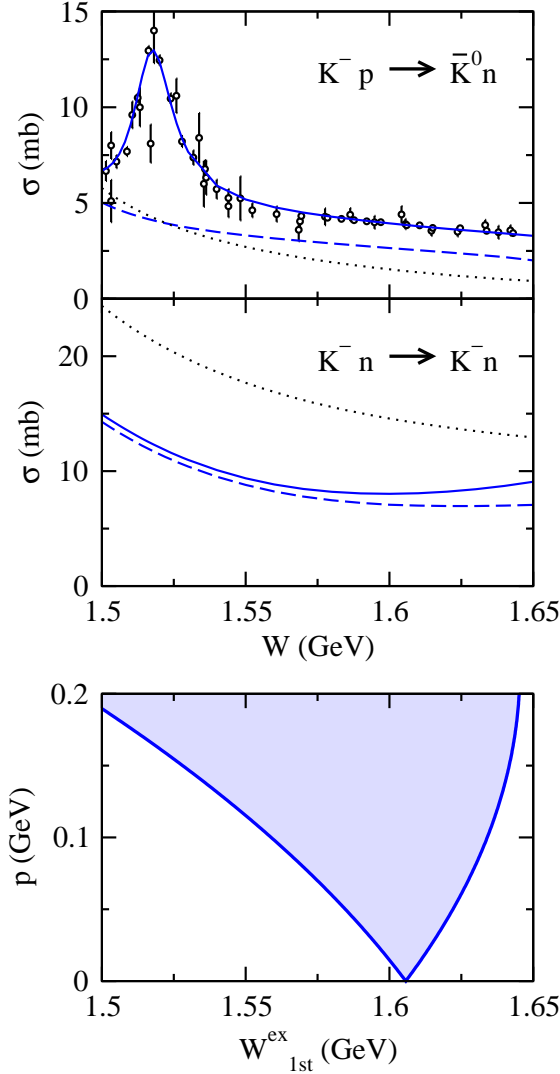}
\caption{
\label{fig:kn-tcs-high-v6}
(Top) Total cross section for $K^- p \to \bar K^0 n$ 
in the energy region relevant to the $\bar K N_2 \to \bar K^{\rm ex} N$ subprocess
in the $\bar K$-exchange process [Fig.~\ref{fig:reaction}(b)]
for the case of $|\vec p_{\bar K}|=0.6$~GeV and $\theta_{p_N} = 0$.
The solid (dashed) curve is the full ($S$-wave only) result from Model B of Ref.~\cite{knlskp1},
while the dotted curve is from the model developed in Ref.~\cite{or}
that was used for the calculation in Refs.~\cite{mh,jido,yamagata}.
(Middle) Same as the top panel but for $K^- n \to \bar K^- n$. 
(Bottom) Allowed ranges of the invariant mass $W^{\rm ex}_{\rm 1st}$
for the $\bar K N_2 \to \bar K^{\rm ex} N$ subprocess as $p\equiv|-\vec p|$ is varied. 
Here the incoming-$\bar K$ momentum and the scattering angle of outgoing $N$ are fixed as
$|\vec p_{\bar K}|=0.6$~GeV and $\theta_{p_N} = 0$.
}
\end{figure}

We next compare our results at  $|\vec p_K|=0.6$ GeV with those given in Ref.~\cite{mh}.
In Fig.~\ref{fig:600mev}, we see that our ``$S$-wave only'' results at $|\vec p_{\bar K}| = 0.6$
GeV are much smaller than the results in Ref.~\cite{mh}.
The results in Ref.~\cite{mh} are even comparable or larger than our full results in which
higher partial waves are also included.
This can be understood from Fig.~\ref{fig:kn-tcs-high-v6}.
For the $\bar K N_2 \to \bar K^{\rm ex} N$ subprocess,
the $K^- p \to \bar K^0 n$ and $K^- n \to K^- n$
charge states can contribute.
We see that at $W\sim 1.6$ GeV, which
corresponds to a typical invariant mass of the $\bar K N_2 \to \bar K^{\rm ex} N$ subprocess
for $|\vec p_{\bar K}| = 0.6$ GeV,
the $S$-wave $\bar K N$ model used in Ref.~\cite{mh}
gives a large cross section for $K^- n \to K^- n$,
which is even larger than our full results.
Because all the $\bar K N$ models give similar cross sections near the threshold,
we can conclude that this is the origin of the large $K^-d \to \pi \Sigma n$
reaction cross section found in Ref.~\cite{mh}.
Furthermore, the $K^-n\rightarrow K^-n$ cross sections are larger than $K^-p \rightarrow K^0 n$
cross sections and thus have a larger contribution to the $\bar K$-exchange amplitudes.
This is why the result from  Ref.~\cite{mh}
has a large cross section for $K^- d\rightarrow \pi \Sigma n$ at $p_K=0.6$ GeV.
This observation also indicates that one must use the $\bar K N$ amplitudes that are well tested by the $\bar K N$ reaction data
up to a high-energy region far beyond the $\bar K N$ threshold.

\section{Summary and future developments}
\label{sec:summary}

Aiming at establishing low-lying $Y^*$ resonances 
through analyzing the forthcoming data from the J-PARC E31 experiment,
we have developed a model for the $\bar K d \to \pi Y N$ reaction.
At the  kinematics of this experiment  that the outgoing nucleon is in
the direction of the incoming $\bar{K}$, the cross sections for this reaction
are dominated by the $\bar{K}$-exchange mechanism. The amplitudes
of this $\bar K$-exchange  process are calculated in our approach by  using
the off-shell amplitudes of $\bar{K}N \to \bar{K}N$ and $\bar{K}N \to \pi Y$
generated from the DCC model developed in Ref.~\cite{knlskp1}.
This DCC model was constructed by fitting the
existing data of $K^- p \to \bar K N, \pi \Sigma, \pi \Lambda, \eta \Lambda, K\Xi$ 
reactions over the wide energy region from the thresholds up to $W=2.1$ GeV.

Most previous works used elementary meson-baryon amplitudes that were constructed
by fitting only to the $\bar K N$ reaction data near the threshold.
However, we have shown that 
if the incoming-$\bar K$ momentum is rather high, as in the case of the J-PARC E31 experiment,
the use of such amplitudes would result in the cross section 
that is order(s) of magnitude 
smaller than the one calculated using the appropriate meson-baryon amplitudes
that reproduce the $\bar K N$ reactions in the energy region far beyond the $\bar K N$ threshold.
This is because the meson-baryon subprocess produced by the reaction 
between the incoming $\bar K$ and the nucleon inside of the deuteron
can have a very high invariant mass, even if the invariant mass of the final $\pi Y$ system
is quite low.

We have shown that the $\bar K d \to \pi Y N$ reactions are useful for
studying low-lying $Y^*$ resonances.
In fact, by comparing the results between our two models, Model~A and Model~B, 
we have found that the behavior of the threefold differential cross sections for $K^- d \to \pi \Sigma n$
[$K^- d \to \pi^- \Lambda p$] below the $\bar K N$ threshold 
are sensitive to the existence and position of the 
$S_{01}$ resonance poles including $\Lambda(1405)1/2^-$
[the $P_{13}$ resonance poles including $\Sigma(1385)3/2^+$].
We have also demonstrated that the $K^- d \to \pi^- \Lambda p$ reaction data would provide useful 
information for judging the existence of an unestablished low-lying $J^P=1/2^+$ $\Sigma$ resonance
with the pole mass $M_R = 1457 -i39$ MeV, which is currently found only in Model B.

Here we note that we have followed the previous works~\cite{mh,jido,yamagata} to 
consider only the impulse and $\bar K$-exchange processes and ignore other
higher-order three-particle final-state interactions.
One possible important correction is the $\pi$-exchange mechanism 
when the invariant mass of the outgoing $\pi N$ state in the final $\pi Y N$ state
is  near the $\Delta$(1232) region.
We have found that it has negligible effects to change our results in the considered special
kinematics shown in Fig.~\ref{fig:kinematics}. 
Nevertheless, our results on the differences between Models A and B should be further quantified by performing
the complete three-particle calculation. This is, however, rather difficult within the framework using
the $\bar K N$ amplitudes of the DCC model of Ref.~\cite{knlskp1} mainly because of the presence
of multi-channel final states, such as $\pi \Lambda N$, $\pi\Sigma N$, $\eta \Lambda N$, and $K \Xi N$,
and of the non separable nature of our meson-baryon amplitudes, which is different from those used in Ref.~\cite{ohnishi},
where the separable nature of the two-body amplitudes was a key to solving the three-body scattering equation.
Clearly, this requires a separated long-term effort.

A necessary and immediate next step towards constructing a reliable $\bar K d$ reaction model
that can be used for the spectroscopic study of low-lying $Y^*$ resonances
would be to include the baryon-exchange processes in addition to $\bar K$- and $\pi$-exchange processes,
so that we can apply our $\bar K d$ reaction model to a wider kinematical region. 
Also, the inclusion of baryon-exchange process would make our model applicable to the study of
$YN$ and $YY$ interactions, 
where the latter is quite interesting in relation to a possible existence of the $H$ dibaryons.
Our investigations in this direction will be presented elsewhere.

\begin{acknowledgments}
The authors would like to thank Dr.~S.~Ohnishi for illuminating discussions on his recent studies of $K^- d \to \pi \Sigma n$.
H.K. would also like to thank Professor~H.~Noumi for useful communications on the status of 
the J-PARC E31 experiment.
This work was supported by Japan Society for the Promotion of Science (JSPS) KAKENHI Grant No.~JP25800149
and by the U.S. Department of Energy, Office of Nuclear Physics Division, under Contract No. DE-AC02-06CH11357.
This research used resources of the National Energy Research Scientific Computing Center
and resources provided on Blues and Fusion,
high-performance computing cluster operated by the Laboratory Computing Resource Center at Argonne National Laboratory. 
\end{acknowledgments}

\end{document}